\documentclass[a4paper]{jpconf}

\usepackage{graphicx}
\usepackage{hyperref}

\newcommand{\dau}{{\rm d+Au}}

\newcommand{\auau}{{\rm Au+Au}}
\newcommand{\cucu}{{\rm Cu+Cu}}
\newcommand{\pp}{{\rm p+p}}
\newcommand{\jpsi}{J/\psi}
\newcommand{\ncoll}{{N_{\rm coll}}}
\newcommand{\rda}{{R_{\rm dAu}}}
\newcommand{\raa}{{R_{\rm AA}}}

\begin{document}
\title{Measurements of Cold Nuclear Matter Effects on $\jpsi$ in the
PHENIX Experiment via Deuteron-Gold Collisions}

\author{Matthew Wysocki, for the PHENIX Collaboration}
\address{Dept. of Physics, University of Colorado, Boulder, CO 80309-0390, USA}
\ead{wysockim@colorado.edu}

\begin{abstract}
A new calculation of $\rda$ has been performed using the 2003
\dau~data and the higher-statistics 2005 \pp~data.  These nuclear
modification factors are compared to calculations using
nuclear-modified PDFs and a $\jpsi$ breakup cross section is
extracted.  These values are then used to project the cold nuclear
matter effects in \auau~collisions.  Additionally, a more data-driven
projection is performed.
\end{abstract}

%
\section{Introduction}
Cold nuclear matter (CNM) effects are those due to being within the
nuclear environment, as opposed to a higher-density medium or vacuum.
These effects are quite interesting on their own, but are even more
important at the Relativistic Heavy Ion Collider (RHIC) as we attempt
to observe the Quark-Gluon Plasma (QGP), a state with much higher
energy density than normal nuclei.  There are several proposed
signatures that could indicate the transition to a QGP, and it is
essential to understand them in the context of what normally occurs in
nuclei.

One such proposed signature is the suppression of $\jpsi$ production
in the deconfined medium of the QGP, due to color-charge screening of
the $c\bar{c}$ interaction~\cite{Matsui:1986dk}.  However, $\jpsi$
production may also be modified due to CNM effects such as nuclear
shadowing, gluon saturation, the EMC effect, or absorption/breakup
within the nuclear remnants.  Therefore it is very important to
quantify these contributions.

%
\section{The PHENIX Experiment}
In order to study CNM effects at RHIC, deuteron and gold ions are
collided at $\sqrt{s_{NN}}$ = 200 GeV.  This was first done in Run-3
and more recently in Run-8, the latter of which is currently being
analyzed.  Since the original analysis of the Run-3
data~\cite{Adler:2005ph}, a new p+p dataset was recorded in Run-5 with
more than an order-of-magnitude increase in $\jpsi$ statistics over
the p+p data from Run-3~\cite{Adare:2006kf}.  In addition, there have
been two years' worth of improvements in the reconstruction software,
the $\jpsi$ signal extraction, and our understanding of the detector.
Consequently, a new analysis of the Run-3 \dau~data was performed
using the same methods as the new \pp~reference data so that an
apples-to-apples comparison could be done~\cite{Adare:2007gn}.  It
should also be noted that the Run-4 \auau~\cite{Adare:2006ns} and
Run-5 \cucu~\cite{Adare:2008sh} $\raa$ results use this same p+p
reference data.

The PHENIX rapidity coverage consists mainly of two regions: the Drift
Chamber, Pad Chamber, EM Calorimeter and RICH measure
$\jpsi\rightarrow e^+e^-$ at mid-rapidity ($|y|<0.35$), and the Muon
Tracker and Muon Identifier measure $\jpsi\rightarrow \mu^+\mu^-$ at
forward and backward rapidity ($1.2<|y|<2.2$).

%
\section{Nuclear Modification Factor}
The nuclear modification factor $\raa$ (Equation~\ref{eq:raa})
quantifies the suppression or enhancement of particle production in
collisions of heavier nuclei with respect to \pp~collisions, scaled by
the appropriate number of binary collisions ($\langle\ncoll\rangle$)
in the heavier species, as calculated by a Glauber
model~\cite{Miller:2007ri}.  Using the Run-3 \dau~and Run-5 \pp~data
we can construct $\rda$ as functions of transverse momentum, collision
centrality, and rapidity.  We will focus on the latter here; for the
broader analysis, see~\cite{Adare:2007gn}.

\begin{equation}\label{eq:raa}
\rda = \frac{1}{\langle \ncoll\rangle} \frac{dN_{\jpsi}^{d+Au} / dy}{dN_{\jpsi}^{p+p} / dy}
\end{equation}

\begin{figure}[thb]
\includegraphics[width=20em]{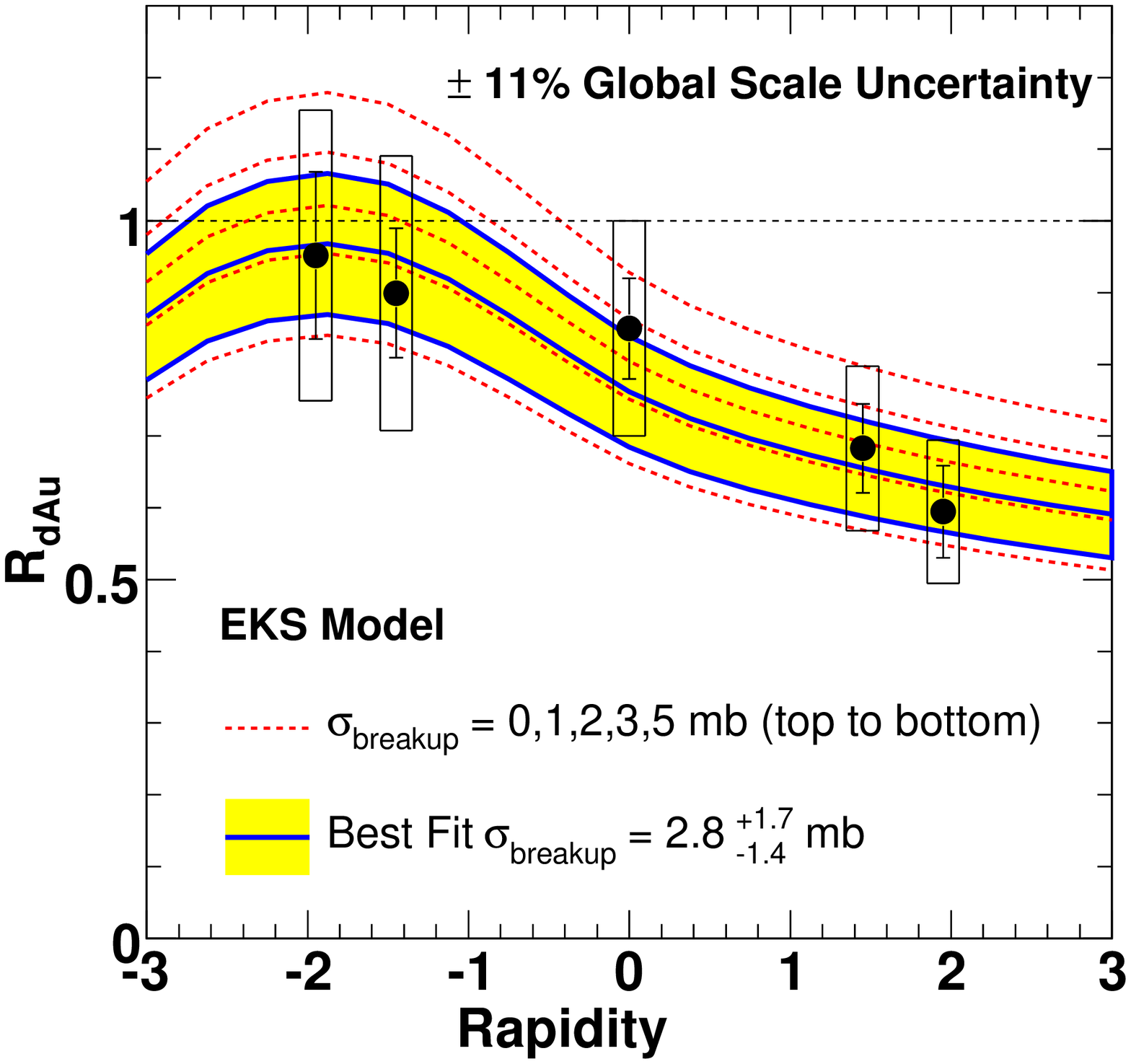} 
\includegraphics[width=20em]{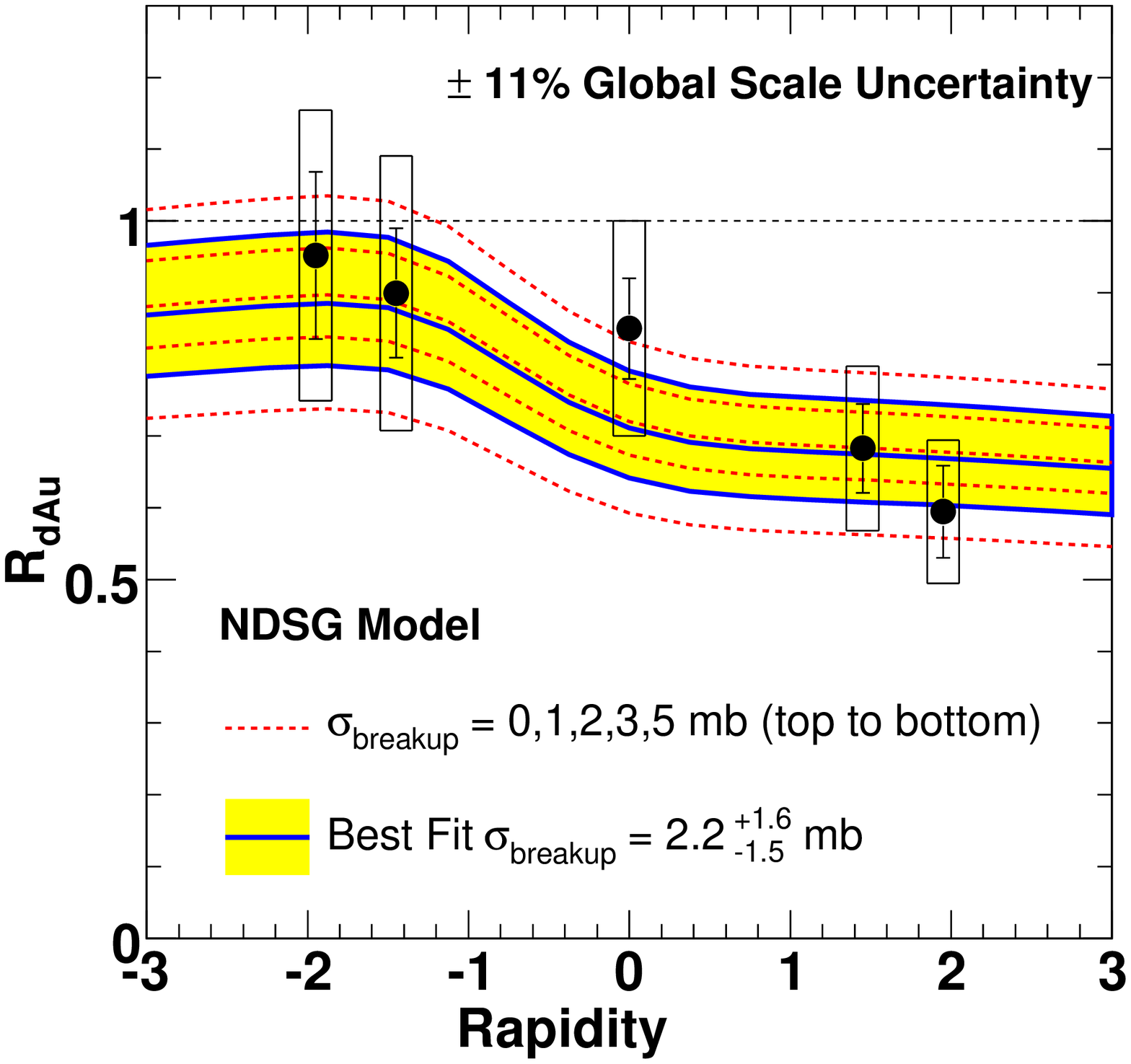} 
\caption{\label{fig:rdau_eksmodel} (color online) $\rda$ data compared
to various theoretical curves for different $\sigma_{breakup}$ values.
Also, shown as a band are the range of $\sigma_{breakup}$ found to be
consistent with the data within one standard deviation.  The left
panel is a comparison for EKS shadowing~\cite{Eskola:2001ek}, while
the right panel is for NDSG shadowing~\cite{NDSG}.}
\end{figure}

$\rda$ as a function of rapidity is plotted as the black points in
Figure~\ref{fig:rdau_eksmodel}.  As can be seen in the Figure, $\jpsi$
production in \dau~collisions is consistent (within the large error
bars) with binary-collision-scaled \pp~collisions at backward and
mid-rapidity, but is suppressed at forward rapidity (in the gold-going
direction).

It can be useful to compare the measured $\rda$ to a simple model of
CNM effects.  By comparing the modification due to available
nuclear-modified PDFs with the data, we can extract a breakup cross
section ($\sigma_{breakup}$) for the $\jpsi$ passing through the
nuclear medium.  This is found to be $\sigma_{breakup} =
2.8^{+1.7}_{-1.4}$ mb and $\sigma_{breakup} = 2.2^{+1.6}_{-1.5}$ mb
using the EKS~\cite{Eskola:2001ek} and NDSG~\cite{NDSG}
nuclear-modified PDFs, respectively.  Note that the quoted
uncertainties account for all experimental statistical and systematic
errors, including the global scale uncertainty.  However, theoretical
model uncertainies are not included.  For details of how the
statistical and systematic errors of the data are accurately accounted
for, see~\cite{Adare:2007gn} and~\cite{Adare:2008cg}.

These values overlap within one standard deviation with the published
value of $\sigma^{\jpsi}_{abs}=4.5\pm0.5$ mb from
CERN-SPS~\cite{Alessandro:2006jt} (within the large uncertainties).
It should be noted, however, that the SPS value does not include any
nuclear shadowing or anti-shadowing.  It has been suggested that the
SPS data lies in the anti-shadowing region, causing an enhancement in
$\jpsi$ production, which would then have to be balanced by an even
larger $\sigma^{\jpsi}_{abs}$ to match the data~\cite{Vogt:2007zzb}.

%
\section{Projections to \auau}
\begin{figure}[htb]
\includegraphics[width=20em]{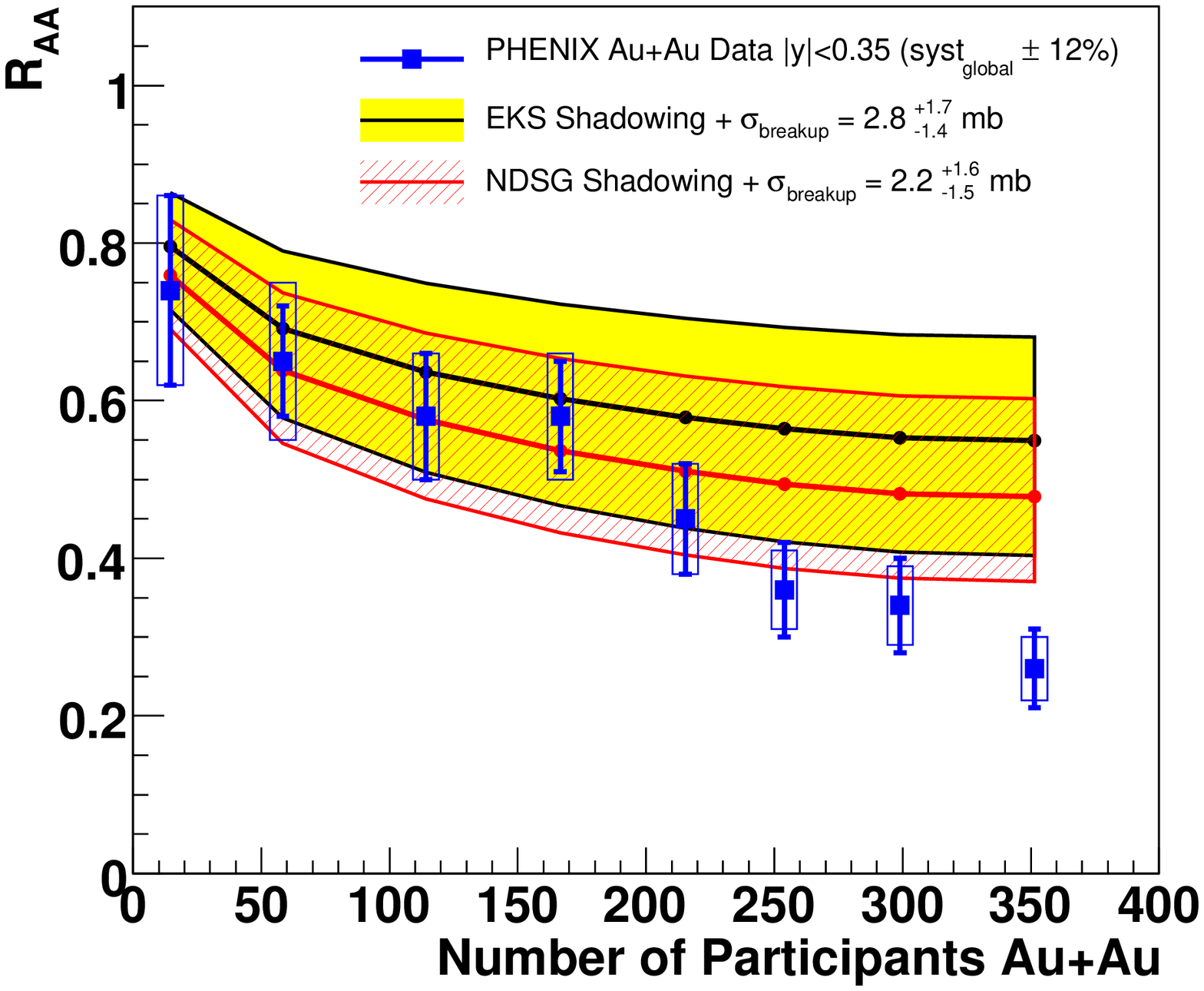}
\includegraphics[width=20em]{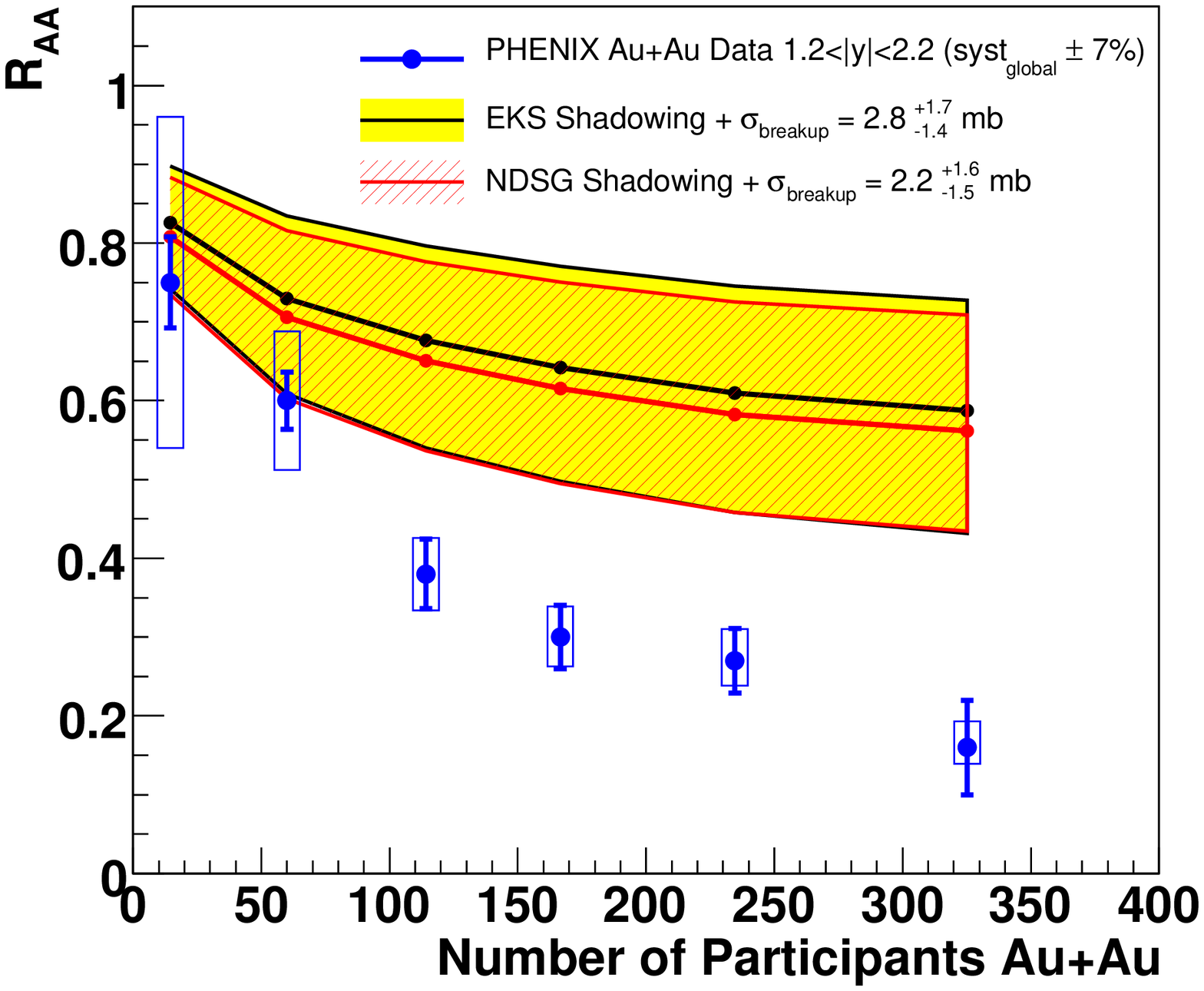}
\caption{\label{fig:rauau_shadowing_projection} (color online) $\raa$
for Au+Au~\cite{Adare:2006ns} collisions compared to a band of
theoretical curves for the $\sigma_{breakup}$ values found to be
consistent with the $\dau$ data as shown in
Figure~\ref{fig:rdau_eksmodel}. The left figure includes both EKS
shadowing~\cite{Eskola:2001ek} and NDSG shadowing~\cite{NDSG} at
mid-rapidity.  The right figure is the same at forward rapidity.}
\end{figure}

To quantify CNM effects in \auau~collisions, we use the
nuclear-modified PDFs (in conjunction with a model of their
impact-parameter dependence~\cite{Klein:2003dj,Vogt:2004dh}) and the
calculated $\sigma_{breakup}$ to project $\raa$ for \auau~strictly
due to CNM effects.  This is shown versus the number of participant
nucleons in the \auau~collision in
Figure~\ref{fig:rauau_shadowing_projection}, where the two bands
represent the one-standard-deviation uncertainty bands using the two
PDFs, and the blue points represent the PHENIX $\jpsi$ $\raa$
measurements from Run-4~\cite{Adare:2006ns}.

The $\raa$ measured in Run-4 shows statistically-significant
suppression at forward rapidity in comparison to the projected
suppression due to CNM effects.  This is not true, however, in the
mid-rapidity case, except perhaps in the most central collisions.
There are several things to note here: first, that the large error
bands on the CNM projections limit our ability to make any definitive
statements, and this will have to be improved in the future; second,
that the above calculations are explicitly model-dependent, and there
is no easy way to include this in the uncertainty bands; third, the
error bands shown in Figure~\ref{fig:rauau_shadowing_projection} are
correlated between the two rapidity regions, as they are due to the
same calculation of $\sigma_{breakup}$.  Calculations in separate
rapidity bins were performed for~\cite{Adare:2007gn}, but are not
included here due to the large uncertainties.

A less model-dependent approach to the problem is to
parametrize the CNM data directly, as proposed
in~\cite{GranierdeCassagnac:2007aj}.  Instead of invoking
nuclear-modified PDFs and breakup cross sections, we start from the
assumption that $\rda$ should approach unity for the most peripheral
collisions, and then fit the measured $\rda$ to a linear function
satisfying this condition (convolving this function with the
\dau~centrality distributions for each data point).  Other fit
functions may be tried, but the results do not vary much due to the
imprecision of the current data.

\begin{figure}[htb]
\includegraphics[width=20em]{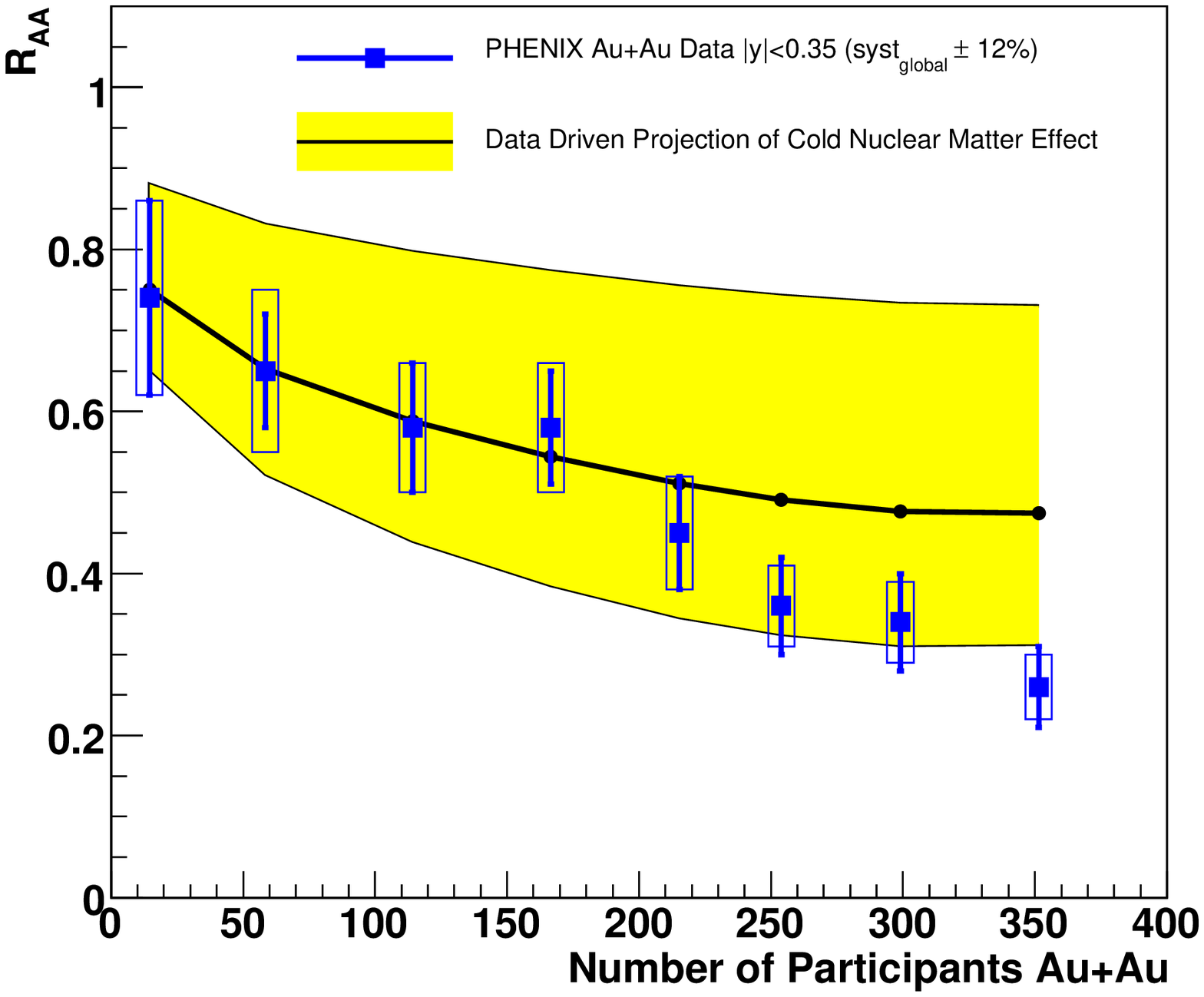}
\includegraphics[width=20em]{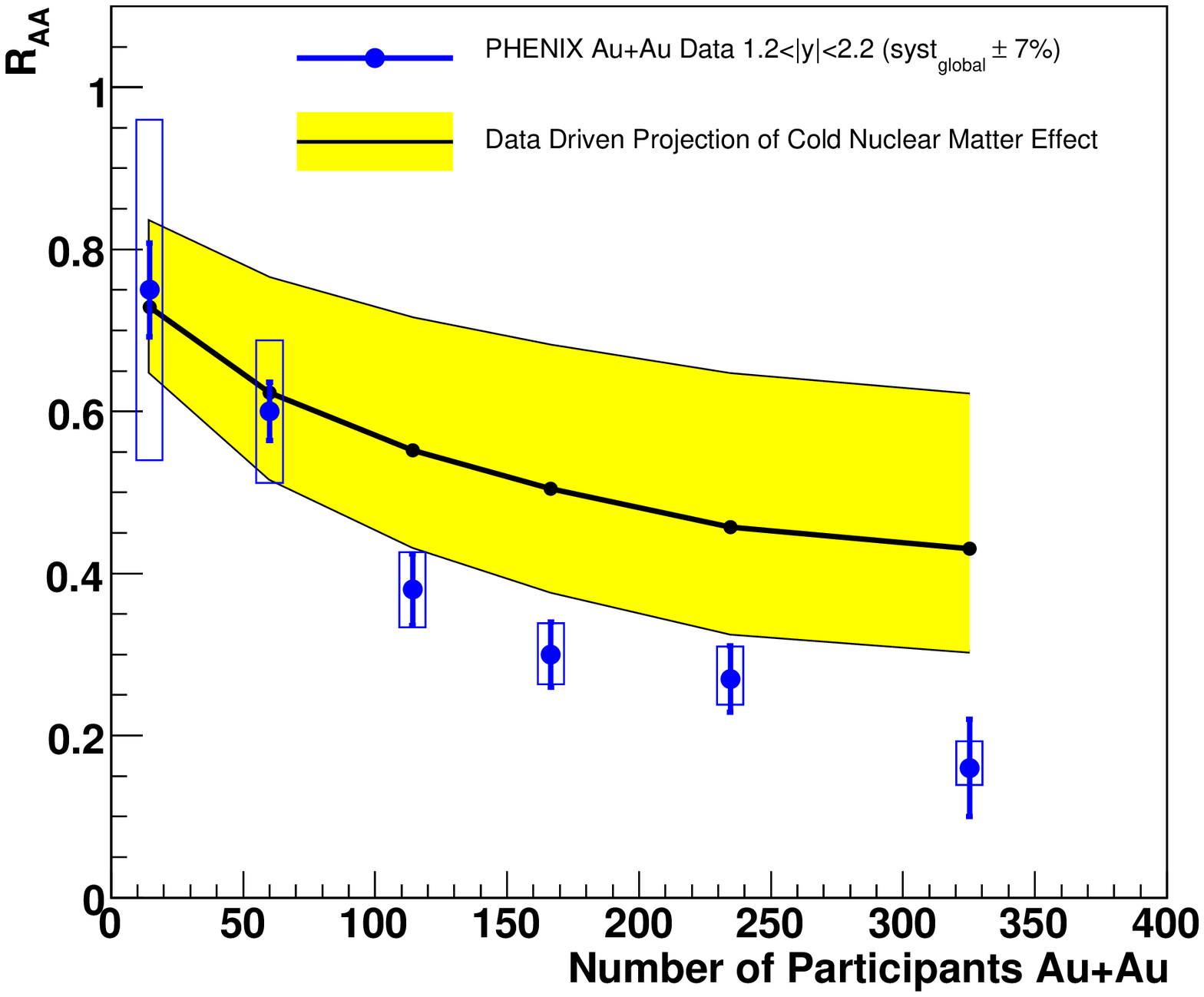}
\caption{\label{fig:rauau_data_projection} Predictions of the
 data-driven method constrained by the $\rda$ as a function of
 collision centrality for the \auau~$\raa$ for mid-rapidity (left) and
 at forward rapidity (right).}
\end{figure}

Using the results of the fit we then construct $\raa$ as a function of
impact parameter by integrating over the radial distribution of binary
collisions for each gold nuclei at a given impact parameter value.  As
can be seen in Figure~\ref{fig:rauau_data_projection}, the resulting
one-standard-deviation bands are qualitatively similar in shape to
those shown in Figure~\ref{fig:rauau_shadowing_projection}, but
generally have a larger $1\sigma$ region and slightly more suppressed
most-likely values.  It must be noted that, contrary to the previous
case, the uncertainty bands here are entirely uncorrelated between
rapidities.

%
\section{Summary}
We have presented $\rda$ as a function of rapidity as a quantification
of CNM effects in heavy ion collisions.  We have also used several
techniques to project the CNM effects to $\raa$ for \auau~collisions.

MGW acknowledges funding from the Division of Nuclear Physics of the
U.S. Department of Energy under Grant No. DE-FG02-00ER41152.


\end{document}